\documentclass[10pt,conference]{IEEEtran}
\IEEEoverridecommandlockouts

\usepackage{cite}
\usepackage{amsmath,amssymb,amsfonts}
\usepackage{algorithmic}
\usepackage{graphicx}
\usepackage{textcomp}
\usepackage{xcolor}

\usepackage{caption}
\usepackage{subcaption}  
\usepackage{enumitem}
\usepackage[english]{babel}
\usepackage{amsthm}

\usepackage{pgfplots}
\usepackage{tikz}
\usetikzlibrary{calc}
\makeatletter
\newcommand{\gettikzxy}[3]{%
  \tikz@scan@one@point\pgfutil@firstofone#1\relax
  \edef#2{\the\pgf@x}%
  \edef#3{\the\pgf@y}%
}
\usetikzlibrary{spy,backgrounds}
\usepackage{pgfplotstable}

\pgfplotsset{compat=newest}
\usetikzlibrary{plotmarks}
\usetikzlibrary{arrows.meta}
\usepgfplotslibrary{patchplots}
\usepackage{grffile}
\newlength\fheight 
\newlength\fwidth 
\usepgfplotslibrary{fillbetween}

\usepackage{acronym}

\acrodef{cc}[CC]{channel charting}
\acrodef{ot}[OT]{optimal transport}
\acrodef{csi}[CSI]{channel state information}
\acrodef{gospa}[GOSPA]{generalized optimal sub-pattern assignment}

\acrodef{5g}[5G]{the fifth generation}
\acrodef{6g}[6G]{the sixth generation}
\acrodef{iot}[IoT]{internet of things}

\acrodef{aoa}[AoA]{angle-of-arrival}
\acrodef{toa}[ToA]{time-of-arrival}
\acrodef{rtt}[RTT]{round-trip time}
\acrodef{tdoa}[TDoA]{time-difference-of-arrival}

\acrodef{ap}[AP]{access point}
\acrodef{bs}[BS]{base station}
\acrodef{cdf}[CDF]{cumulative distribution function}
\acrodef{crb}[CRB]{Cram\'er-Rao bound}
\acrodef{gnss}[GNSS]{global navigation satellite system}
\acrodef{los}[LoS]{line-of-sight}
\acrodef{nlos}[NLoS]{non-line-of-sight}
\acrodef{mae}[MAE]{mean absolute value}
\acrodef{rmse}[RMSE]{root mean square error}
\acrodef{ml}[ML]{machine learning}
\acrodef{mpc}[MPC]{multipath component}
\acrodef{music}[MUSIC]{multiple signal classification}
\acrodef{ofdm}[OFDM]{orthogonal frequency division multiplexing}
\acrodef{snr}[SNR]{signal-to-noise ratio}

\acrodef{ue}[UE]{user equipment}
\acrodef{upa}[UPA]{uniform planar array}
\acrodef{ula}[ULA]{uniform linear array}

\acrodef{mimo}[MIMO]{multiple-input-multiple-output}
\acrodef{dmimo}[D-MIMO]{distributed MIMO}

\acrodef{ml}[ML]{machine learning}
\acrodef{nn}[NN]{neural network}
\acrodef{mlp}[MLP]{multi-layer perceptron}
\acrodef{cnn}[CNN]{convolutional neural network}

\acrodef{music}[MUSIC]{multiple signal classification}
\acrodef{omp}[OMP]{orthogonal matching pursuit}
\acrodef{mae}[MAE]{mean absolute error}
\acrodef{cdf}[CDF]{cumulative distribution function}

\definecolor{myColor}{RGB}{0, 100, 0}

\usepackage{titlesec}
\titlespacing{\section}{0pt}{4pt}{2pt} 
\titlespacing{\subsection}{0pt}{4pt}{2pt}

\usepackage{geometry}
\begin{document}

\IEEEaftertitletext{\vspace{-5mm}}


\title{Localization in Multi-Panel Massive MIMO With Clock Asynchronism: A Unified Approach
\thanks{This work was supported in part by the European Union through the project ISLANDS (grant agreement no. 101120544), and the Swedish Research Council (VR) through the project 6G-PERCEF (Grant 2024-04390) (e-mail: \{yuhaozh, guangjin.pan, furkan, henkw\}@chalmers.se; \{carmen.dandrea, buzzi\}@unicas.it).}
}


\author{\IEEEauthorblockN{Yuhao Zhang\IEEEauthorrefmark{1}, 
Guangjin Pan\IEEEauthorrefmark{1}, 
Musa Furkan Keskin\IEEEauthorrefmark{1}, 
Carmen D'Andrea\IEEEauthorrefmark{2}\IEEEauthorrefmark{3}, 
Stefano Buzzi\IEEEauthorrefmark{2}\IEEEauthorrefmark{3}, 
Henk Wymeersch\IEEEauthorrefmark{1}}
\IEEEauthorblockA{\IEEEauthorrefmark{1}Department of Electrical Engineering, Chalmers University of Technology, Sweden\\
\IEEEauthorrefmark{2}Department of Electrical and Information Engineering, University of Cassino and Southern Lazio, Italy\\
\IEEEauthorrefmark{3}Consorzio Nazionale Interuniversitario per le Telecomunicazioni (CNIT), Italy}}

\maketitle

\begin{abstract}
    In this work, we propose a unified localization framework, termed UNILocMP, that combines model-based geometry and \ac{cc} for multi-panel massive \ac{mimo} under clock asynchronism. Owing to the multi-panel architecture, users are classified into multi-\ac{los} users, which maintain \ac{los} links with at least two panels, and single/non-\ac{los} users, which maintain a \ac{los} link with only one panel or with none of the panels. For multi-\ac{los} users, a joint position and clock bias estimation is developed; while for single/non-\ac{los} users, an unsupervised \ac{cc} model is trained with a two-stage data augmentation strategy, where a clock-bias aware dissimilarity metric is introduced. It is numerically validated that the proposed UNILocMP outperforms model-based and \ac{cc}-based baselines and achieves acceptable performance compared with fully-supervised fingerprinting. Moreover, for a fixed total number of antennas, the multi-panel architecture significantly improves localization accuracy and robustness compared with a single-panel \ac{bs} deployment, particularly in the presence of clock asynchronism.
\end{abstract}


\acresetall

\section{Introduction}
The evolution toward future wireless systems attracts significant attention to high-precision localization and sensing, enabling a wide range of applications such as navigation and autonomous driving, intelligent transportation systems, smart factories, and immersive extended reality~\cite{Near_Loca_Wang_2025,Tran_Coll_Xu_2026}. Recently, multi-panel massive \ac{mimo} has emerged as a promising architecture due to their capability to improve spatial coverage and geometric diversity~\cite{Mult_MIMO_Huan_2018,Join_Acti_Xiu_2023}. Compared with conventional single-\ac{bs} arrays, multi-panel \ac{mimo} deployments distribute multiple antenna panels over a larger physical area, thereby providing enhanced \ac{los} opportunities, fine spatial resolution, and multiple geometric perspectives. By leveraging \ac{csi} collected from multiple distributed panels, localization can potentially achieve substantially improved accuracy and robustness~\cite{5G_Mult_Gert_2020}. 

Localization methods can generally be divided into two categories: model-based and data-driven methods~\cite{5G_Mult_Gert_2020,A_Tuto_Chen_2022,Angl_Prof_Step_2024,Tran_Coll_Xu_2026}. Given complementary strengths and weaknesses, these two methods can be integrated in a mutually beneficial manner, and some works have attempted to combine model-based localization and \ac{cc} to improve positioning performance~\cite{Glob_Scal_Omid_2024,TDoA_Self_Mohs_2025,Augm_Chan_Euch_2023,Angl_Prof_Step_2024,Unif_Pro_Yuha_2025}. However, most existing works either require additional sensing hardware~\cite{Glob_Scal_Omid_2024,TDoA_Self_Mohs_2025} or rely on \ac{los} dominant environments~\cite{TDoA_Self_Mohs_2025,Augm_Chan_Euch_2023,Angl_Prof_Step_2024}. In mixed \ac{los}/\ac{nlos} scenarios, a unified localization framework, i.e., UNILocPro, is proposed in~\cite{Unif_Pro_Yuha_2025} for single-\ac{bs} localization, which achieves comparable performance to fully-supervised fingerprinting without requiring labeled data. However, UNILocPro is designed for single-\ac{bs} settings and requires tight synchronization between the user and \ac{bs}. In practice, maintaining tight synchronization is challenging and costly, and residual clock biases inevitably remain due to, e.g., oscillator drift and hardware impairments~\cite{Enha_Fing_Xu_2026}.

In the literature, the integration of model-based geometry and data-driven learning for multi-panel localization is underexplored, especially under mixed \ac{los}/\ac{nlos} conditions. In multi-panel systems, users experience heterogeneous and space-varying visibility across different panels, which necessitates seamless and visibility-aware integration. In this work, we aim for a unified localization framework for spatially distributed multi-panel massive \ac{mimo} systems with clock asynchronism in mixed \ac{los}/\ac{nlos} environments. The main contributions are summarized as:
\begin{itemize}
\item We extend the unified localization philosophy of UNILocPro to multi-panel systems and propose a visibility-aware framework, i.e., UNILocMP, that integrates model-based geometry and \ac{cc} to address heterogeneous visibility conditions while exploiting the geometric diversity of distributed panels.\footnote{UNILocPro considers a single-\ac{bs} deployment with tight synchronization, whereas UNILocMP considers a spatially distributed multi-panel architecture with residual clock asynchronism and heterogeneous \ac{los} visibility. While the training principles inherited from UNILocPro serve as the learning foundation, the multi-panel visibility modeling, joint position and clock bias estimation, multi-panel dissimilarity, and clock-bias-aware adaptation constitute the main contributions developed in this work.}
\item A joint position and clock bias optimization method is proposed to directly match the observed \ac{csi} with the geometry-induced \ac{los} channel across visible panels. A closed-form \ac{tdoa}/\ac{aoa} multi-lateration is further derived to provide an efficient initialization for the subsequent optimization. Moreover, we adapt \ac{cc} to the multi-panel and asynchronous setting by introducing a multi-panel fused, clock-bias-aware dissimilarity metric and a clock-bias-oriented \ac{csi} data augmentation strategy.
\item Ray-tracing simulations are conducted to evaluate single-\ac{bs} and multi-panel localization under both perfect and imperfect synchronization. The results demonstrate that the proposed framework achieves improved performance by exploiting the complementary spatial information from multiple panels.
\end{itemize}

\section{System Model}

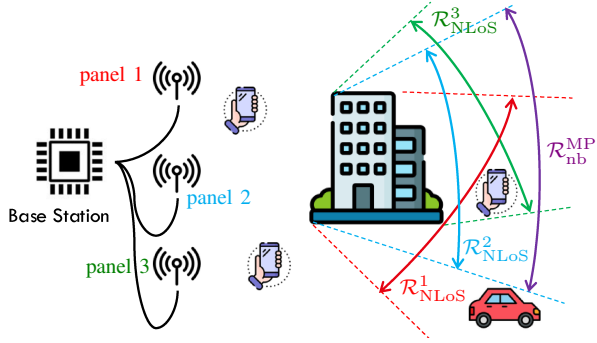
\begin{figure}[t]
    \centering
    \begin{tikzpicture}[every node/.style={font=\footnotesize}]
    \node (image) [anchor=south west]{\includegraphics[width=0.9\linewidth]{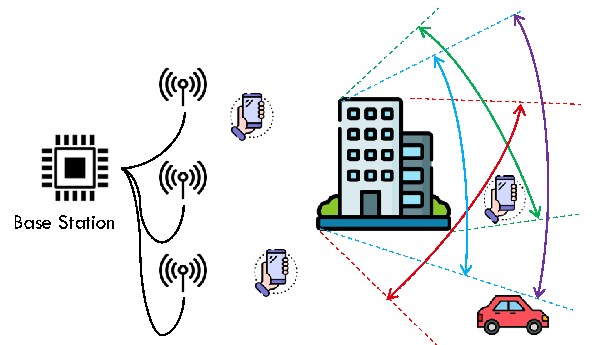}};
    \gettikzxy{(image.north east)}{\ix}{\iy};
    
    \node at (0.71*\ix,0.22*\iy)[rotate=0,anchor=north]{\textcolor{red}{$\mathcal{R}_{\rm NLoS}^1$}};
    \node at (0.81*\ix,0.33*\iy)[rotate=0,anchor=north]{\textcolor{cyan}{$\mathcal{R}_{\rm NLoS}^2$}};
    \node at (0.76*\ix,0.96*\iy)[rotate=0,anchor=north]{\textcolor{green!50!black}{$\mathcal{R}_{\rm NLoS}^3$}};
    \node at (0.93*\ix,0.6*\iy)[rotate=0,anchor=north]{\textcolor{violet}{$\mathcal{R}_{\mathrm{nb}}^{\mathrm{MP}}$}};

    \node at (0.2*\ix,0.8*\iy)[rotate=0,anchor=north]{\textcolor{red}{panel 1}};
    \node at (0.37*\ix,0.46*\iy)[rotate=0,anchor=north]{\textcolor{cyan}{panel 2}};
    \node at (0.21*\ix,0.27*\iy)[rotate=0,anchor=north]{\textcolor{green!50!black}{panel 3}};

    \end{tikzpicture}
    \caption{Illustration of a multi-panel massive \ac{mimo} system in a street-canyon scenario, where colored regions indicate \ac{los} coverage of each panel.}
    \label{fig:system_illustration}
    \vspace{-3mm}
\end{figure}

We consider a multi-panel massive \ac{mimo} system, as depicted in Fig.~\ref{fig:system_illustration}, where a \ac{bs} is equipped with $P$ spatially separated antenna panels to serve multiple single-antenna users, located at positions $\mathbf{p}_i = [ x_i, y_i, z_i]^\mathsf{T}$, $\forall i$, over \ac{ofdm} waveform with $N_c$ subcarriers (the subcarrier spacing is $\Delta_f$). It is assumed that each panel $p \in \{1, \ldots, P\}$ deploys an \ac{ula} with $M_p$ antennas (the total number of antennas deployed by the \ac{bs} is $M = \sum_p M_p$), and the $P$ panels are widely distributed, e.g., across a building facade or within a street canyon.\footnote{Note that the height of each user, i.e., $z_i$, is assumed to be known in this work such that \ac{ula} is sufficient for localization; otherwise, a \ac{upa} can be used to provide more spatial measurements.} The position of panel $p$ is denoted by $\mathbf{p}_\mathrm{panel}^p = [ x_p, y_p, z_p]^\mathsf{T}$. We assume that the \ac{bs} can synchronize its panels with each other, which can be achieved by, e.g., using a common reference clock~\cite{BDS_GPS_Song_2025}. However, due to various factors, users are not perfectly synchronized with the \ac{bs}, which leads to residual clock biases, which for a given user $i$ can be characterized as an unknown value within a bounded interval, denoted by $\epsilon_i \in [-\epsilon_{\max}, \epsilon_{\max}]$.




\subsection{Channel Model}\label{sec:channel_model}
Due to the large separation and aperture of the multi-panel array, the channel of each user exhibits spatial non-stationarity~\cite{Mult_MIMO_Huan_2018}. For example, a user may only be visible, i.e., possess a \ac{los} path, to a subset of panels. The frequency domain channel matrix between user $i$ and panel $p$, denoted by $\mathbf{H}_i^p \in \mathbb{C}^{M_p \times N_c}$, can be modeled as a sum of multi-path components $\mathbf{H}_i^p = \sum_{l = 1}^{L_i^p} \beta_{l,i}^p \mathbf{a}_p(\theta_{l,i}^p) \mathbf{b}^\mathsf{T}(\tau_{l,i}^p + \epsilon_i)$~\cite{Mode_End_Jose_2025,Fund_Trad_Kesk_2025}, where there are $L_i^p$ channel paths from user $i$ to panel $p$; $\beta_{l,i}^p$, $\theta_{l,i}^p$, and $\tau_{l,i}^p$ are the complex channel gain, \ac{aoa}, and \ac{toa} of the $l$-th path, respectively; $\mathbf{b}(\tau)$ is the frequency-domain steering vector; $\mathbf{a}_p(\theta)$ is the array steering vector for panel $p$. The aggregated \ac{csi} for user $i$ across all panels and subcarriers is denoted by $\boldsymbol{\mathcal{H}}_i = \{{ \mathbf{H}_i^p }\}_p$.


Focusing on clock asynchronism and to avoid conflating \ac{csi} estimation errors with localization methods, we assume that the system operates in a high \ac{snr} regime, where the \ac{csi} can be realiably estimated at the \ac{bs}.\footnote{As illustrated in~\cite{Unif_Pro_Yuha_2025}, while \ac{csi} estimation errors would affect the localization accuracy for all methods, the relative performance trend among different methods remains consistent.} With \ac{ofdm} transmission and multi-antenna array, channel parameters at each panel, including channel gain, \ac{aoa}, and \ac{toa} for different channel paths, can be estimated from the \ac{csi} $\mathbf{H}_i^p$ by, for example, the \ac{omp} algorithm~\cite{Mode_End_Jose_2025}. Specifically, with \ac{omp} on the \ac{csi} $\mathbf{H}_i^p$, $\hat{L}_i^p$ channel paths can be recovered with channel parameters $\{ \hat{\beta}_{l,i}^p, \hat{\theta}_{l,i}^p, \hat{\tau}_{l,i}^p \}_{l = 1,2,\ldots,\hat{L}_i^p}$.



\subsection{Map Information}\label{sec:map_information}
We assume the availability of a coarse map of the environment $\boldsymbol{\mathcal{M}}$, from which the region for all possible user positions can be extracted as $\mathcal{R} \subseteq \mathbb{R}^{3 \times 1}$, i.e., $\mathbf{p}_i \in \mathcal{R}$, $\forall i$.\footnote{The map is not assumed to provide high-resolution geometric information; it is used primarily to provide coarse feasible-region and learning constraints.} By validating whether a given point in $\mathcal{R}$ has a \ac{los} connection to the panel $p$, the map $\boldsymbol{\mathcal{M}}$ allows us to partition $\mathcal{R}$ into \ac{los} and \ac{nlos} regions relative to each panel $p$, denoted as $\mathcal{R}_{\mathrm{LoS}}^p \subseteq \mathbb{R}^{3 \times 1}$ and $\mathcal{R}_{\mathrm{NLoS}}^p \subseteq \mathbb{R}^{3 \times 1}$, as shown in Fig.~\ref{fig:system_illustration}. Building on these partitions relative to individual panels, we can define two user position categories: 1) $\mathbf{p}_i$ is considered a \textbf{\emph{multi-\ac{los} user position}} if it has \ac{los} connections to at least two panels, i.e., $\mathbf{p}_i \in \mathcal{R}_{\mathrm{mb}}^{\mathrm{MP}} = \cup_{p \neq p'} (\mathcal{R}_{\mathrm{LoS}}^p \cap \mathcal{R}_{\mathrm{LoS}}^{p'})$; 2) $\mathbf{p}_i$ is considered a \textbf{\emph{single/non-\ac{los} user position}} if it has a LoS connection with only one panel or with none of the panels, i.e., $\mathbf{p}_i \in \mathcal{R}_{\mathrm{nb}}^{\mathrm{MP}} = \mathcal{R} \setminus \mathcal{R}_{\mathrm{mb}}^{\mathrm{MP}}$.

\section{Multi-Panel Unified Localization}\label{sec:UNILocMP}
In this paper, the goal of localization is to estimate the position of each user $\mathbf{p}_i$, as denoted by $\hat{\mathbf{p}}_i$, based solely on $\boldsymbol{\mathcal{H}}_i$. In UNILocMP, user identification is first conducted based on \ac{csi} $\boldsymbol{\mathcal{H}}_i$ from an identification function $I_\mathrm{MP}(\cdot)$, where $I_\mathrm{MP}(\boldsymbol{\mathcal{H}}_i)=1$ indicates that user $i$ is identified as a multi-\ac{los} user; otherwise, user $i$ is identified as a single/non-\ac{los} user; if $I_\mathrm{MP}(\boldsymbol{\mathcal{H}}_i)=1$, the model-based method would be used; otherwise, a \ac{nn} model is adopted:
\begin{equation*}
    \hat{\mathbf{p}}_i = \left\{ \begin{array}{ll}
        f_{\rm me}(\boldsymbol{\mathcal{H}}_i) , & {\rm if} \, I_\mathrm{MP}(\boldsymbol{\mathcal{H}}_i) = 1;  \\
        f_{\Theta}\left[f_{\rm extr}^\mathrm{C}(\boldsymbol{\mathcal{H}}_i)\right], & {\rm if} \, I_\mathrm{MP}(\boldsymbol{\mathcal{H}}_i) = 0,
    \end{array} \right.
\end{equation*}
where $f_{\rm me}(\cdot)$ is a model-based estimation function; $f_{\rm extr}^\mathrm{C}(\cdot)$ is a feature extraction function to transform \ac{csi} $\boldsymbol{\mathcal{H}}_i$ into the input of the \ac{nn} model; $f_{\Theta}(\cdot)$ is the forward function of the \ac{nn} model with $\Theta$ as parameters. 

\subsection{CSI-Based User Identification}\label{sec:User_Identification}
It is assumed that we have a per-panel \ac{csi}-based \ac{los}/\ac{nlos} identification mechanism~\cite{An_Intr_Zeka_2012,Unif_Pro_Yuha_2025}, denoted by $I(\cdot)$, where $I(\mathbf{H}_i^p)=1$ indicates that user $i$ has a \ac{los} path to panel $p$, and otherwise when $I(\mathbf{H}_i^p)=0$.  In the literature, various identification mechanisms have been proposed, achieving up to $93\% \text{-} 100\%$ accuracy~\cite{An_Intr_Zeka_2012}; therefore, instead of designing a specific mechanism, we treat $I(\cdot)$ as external and denote its identification accuracy as $p_\mathrm{I} \in [0.5, 1]$. According to the definition of user positions in Sec.~\ref{sec:map_information}, user identification can be conducted on a per-panel basis, and the multi-panel identification function $I_\mathrm{MP}(\cdot)$ can be expressed as
\begin{equation}\label{eq:multi_LoS_identification_system}
    I_\mathrm{MP}(\boldsymbol{\mathcal{H}}_i) = \left\{ \begin{array}{ll}
        1 , & {\rm if} \, \sum_p I(\mathbf{H}_i^p) \geq 2,  \\
        0 , & {\rm if} \, \sum_p I(\mathbf{H}_i^p) \leq 1,
    \end{array} \right. \quad \forall i.
\end{equation}

\subsection{Model-Based Multi-Panel Localization}\label{sec:model_based_localization}
Once user $i$ is identified as a multi-\ac{los} user, the model-based localization function $f_{\rm me}(\cdot)$ is invoked. Unlike some existing multi-lateration methods that minimize discrepancies between extracted measurements (e.g., \ac{aoa}, \ac{toa}, or \ac{tdoa}) and their geometric predictions~\cite{5G_Mult_Gert_2020,A_Tuto_Chen_2022}, we develop a joint estimation method that directly optimizes in the \ac{csi} domain by matching the measured \ac{csi} with the analytical channel model parameterized by the user position and clock bias. Moreover, we incorporate map-derived visibility constraints to ensure geometric consistency across panels and exclude infeasible positions. Specifically, let $\mathcal{P}_{\rm LoS}(i) = \{ p \in \{1, \ldots, P\}: I(\mathbf{H}_i^p) =1\}$ denote the set of panels identified as having a \ac{los} path to user $i$. The position estimation problem can be formulated as finding a candidate position $\mathbf{p}_{\rm c}^i = [x_i, y_i, z_i]^\mathsf{T} \in \cap_{p \in \mathcal{P}_{\rm LoS}(i)} \mathcal{R}_{\mathrm{LoS}}^{p}$ whose associated \ac{los} channels from all \ac{los} panels best explain the \ac{csi} $\boldsymbol{\mathcal{H}}_i$. For each panel $p \in \mathcal{P}_{\rm LoS}(i)$, the geometric parameters can be computed:
\begin{align}
    \tau^p(\mathbf{p}_{\rm c}^i) &= \| \mathbf{p}_\mathrm{panel}^p - \mathbf{p}_{\rm c}^i \|_2 / c, \label{eq:geometric_relationships_ULA_1} \\
    \sin \left( \theta^p(\mathbf{p}_{\rm c}^i) \right) &= (\mathbf{p}_\mathrm{panel}^p - \mathbf{p}_{\rm c}^i)^\mathsf{T} \mathbf{n}_{\rm ULA}^p / \| \mathbf{p}_\mathrm{panel}^p - \mathbf{p}_{\rm c}^i \|_2, \label{eq:geometric_relationships_ULA_2}
\end{align}
where $\mathbf{n}_{\rm ULA}^p$ is the normalized vector representing the direction of the \ac{ula} for panel $p$ and $c$ is the speed of light. Therefore, the panel-specific \ac{los} channel from $\mathbf{p}_{\rm c}^i$ to the panel $p \in \mathcal{P}_{\rm LoS}(i)$ can be written as $\mathbf{H}^p_\mathrm{LoS} (\mathbf{p}_{\rm c}^i, \epsilon_i) = \beta_\mathrm{LoS}^p (\mathbf{p}_{\rm c}^i) \mathbf{a}_p(\theta^p(\mathbf{p}_{\rm c}^i)) \mathbf{b}^\mathsf{T}(\tau^p(\mathbf{p}_{\rm c}^i) + \epsilon_i)$, where $\beta_\mathrm{LoS}^p (\mathbf{p}_{\rm c}^i)$ is the (unknown) complex \ac{los} channel gain.

\subsubsection{Joint Optimization}
The position optimization problem can be formulated by minimizing the discrepancy between the observed \ac{csi} $\boldsymbol{\mathcal{H}}_i$ and the geometrically predicted \ac{los} channels across all \ac{los} panels. However, this joint estimation assumes that the \ac{los} path is dominant, which may not hold in scenarios with strong \ac{nlos} components. In this work, given estimated channel parameters $\{ \hat{\beta}_{l,i}^p, \hat{\theta}_{l,i}^p, \hat{\tau}_{l,i}^p \}_{l = 1,2,\ldots,\hat{L}_i^p}$ for each panel $p$, we filter out strong \ac{nlos} paths by excluding the paths whose channel gain is above the gain of the shortest path, based on which the remaining \ac{csi} can be written as $\tilde{\mathbf{H}}_i^p = \mathbf{H}_i^p - \sum_{l \in \mathcal{L}_{\mathrm{r}}^{i,p}} \hat{\beta}_{l,i}^p \mathbf{a}_p(\hat{\theta}_{l,i}^p) \mathbf{b}^\mathsf{T}(\hat{\tau}_{l,i}^p)$, where $\mathcal{L}_{\mathrm{r}}^{i,p} = \{ l: |\hat{\beta}_{l,i}^p| > |\hat{\beta}_{l_i',i}^p|, \, l \neq l_i' \}$ with $l_i' = \arg \min_l \hat{\tau}_{l,i}^p$ being the index of the shortest path for user $i$. Consequently, the squared Frobenius norm error can be written as $\sum_{p \in \mathcal{P}_{\rm LoS}(i)} \| \tilde{\mathbf{H}}_i^p - \mathbf{H}^p_\mathrm{LoS} (\mathbf{p}_{\rm c}^i,\epsilon_i) \|_F^2$, which can be minimized over $\mathbf{p}_{\rm c}^i$, $\epsilon_i$, and $\{\beta_\mathrm{LoS}^p\}_{p \in \mathcal{P}_{\rm LoS}(i)}$ jointly. Furthermore, for given $\mathbf{p}_{\rm c}^i$ and $\epsilon_i$, the optimal $\beta_\mathrm{LoS}^p$ for each panel $p$ can be obtained by solving a least-squares problem, based on which a reduced problem with a concentrated cost function~\cite{Maxi_Like_Swin_1998} can be formulated as
\begin{equation}\label{eq:reduced_opt_problem}
    \begin{aligned}
    \max_{\mathbf{p}_{\rm c}^i, \, \epsilon_i}  \quad &
    \mathcal{J}(\mathbf{p}_{\rm c}^i, \epsilon_i), \\
    \mathrm{s.t.} \quad & \mathbf{p}_{\rm c}^i \in \cap_{p \in \mathcal{P}_{\rm LoS}(i)} \mathcal{R}_{\mathrm{LoS}}^{p}, \, -\epsilon_{\max} \leq \epsilon_i \leq \epsilon_{\max},
\end{aligned} 
\end{equation}
where
\begin{equation*}
    \mathcal{J}(\mathbf{p}_{\rm c}^i, \epsilon_i)  =  \!\!\!\!\!\!\!\! \sum_{p \in \mathcal{P}_{\rm LoS}(i)} \!\! \frac{1}{M_p} | \langle \tilde{\mathbf{H}}_i^p, \, \mathbf{a}_p(\theta^p(\mathbf{p}_{\rm c}^i)) \mathbf{b}^\mathsf{T}(\tau^p(\mathbf{p}_{\rm c}^i)+\epsilon_i) \rangle_F |^2.
\end{equation*}


It can be seen that the problem~\eqref{eq:reduced_opt_problem} is non-convex since the objective $\mathcal{J}(\mathbf{p}_{\rm c}^i,\epsilon_i)$ is non-convex w.r.t. $\mathbf{p}_{\rm c}^i$ and $\epsilon_i$, and the feasible region $\cap_{p \in \mathcal{P}_{\rm LoS}(i)} \mathcal{R}_{\mathrm{LoS}}^{p}$ is typically non-convex and possibly disconnected in complex environments. Although grid search over position and clock bias can be adopted to find an optimum, its computational complexity is prohibitive. Therefore, we adopt a gradient-based optimization starting from an initial point $(\mathbf{p}_{\rm c}^{i,0}, \epsilon_i^0)$ to find a local optimum of~\eqref{eq:reduced_opt_problem}:
\begin{equation}\label{eq:gradient_based_optimization}
    \max_{\mathbf{p}_{\rm c}^i, \, \epsilon_i}  \quad 
    \mathcal{J}(\mathbf{p}_{\rm c}^i, \epsilon_i), \quad \mathrm{s.t.} \quad \mathbf{p}_{\rm c}^i \in \mathcal{R}(\mathbf{p}_{\rm c}^{i,0}), \, \epsilon_i \in \mathcal{E}(\epsilon_i^0),
\end{equation}
where $\mathcal{R}(\mathbf{p}_{\rm c}^{i,0})$ (resp. $\mathcal{E}(\epsilon_i^0) \subseteq [-\epsilon_{\max}, \epsilon_{\max}]$) is a convex neighborhood of $\mathbf{p}_{\rm c}^{i,0}$ (resp. $\epsilon_i^0$). Note that the initial point should be chosen properly to ensure that the local optimum found by~\eqref{eq:gradient_based_optimization} is close to the optimum of~\eqref{eq:reduced_opt_problem}.

\subsubsection{Initial Point Acquisition}
The initial point $(\mathbf{p}_{\rm c}^{i,0}, \epsilon_i^0)$ is obtained by multi-lateration based on \ac{tdoa} and \ac{aoa} from all \ac{los} panels. Specifically, given the estimated $\{ \hat{\beta}_{l,i}^p, \hat{\theta}_{l,i}^p, \hat{\tau}_{l,i}^p \}_l$, a reference panel $p_\mathrm{r}$ is set as the panel with the highest overall channel gain, and the \ac{tdoa} is obtained by first selecting the \ac{toa} of the shortest path for each panel $\hat{\tau}_{\mathrm{S},i}^p = \min_l \hat{\tau}_{l,i}^p$ ($\hat{\theta}_{\mathrm{S},i}^p$ is the associated \ac{aoa} estimate), and then taking the difference against the reference panel, i.e., $\hat{\tau}_{\mathrm{TDoA},i}^p = \hat{\tau}_{\mathrm{S},i}^p - \hat{\tau}_{\mathrm{S},i}^{p_\mathrm{r}}$, $\forall p \in \mathcal{P}_{\rm LoS}(i)$. Together with~\eqref{eq:geometric_relationships_ULA_2}, a system of geometric equations can be established:
\begin{align}
    & c \cdot \hat{\tau}_{\mathrm{TDoA},i}^p = \| \mathbf{p}_\mathrm{panel}^p - \mathbf{p}_{\rm c}^{i, 0} \|_2 - \| \mathbf{p}_\mathrm{panel}^{p_\mathrm{r}} - \mathbf{p}_{\rm c}^{i, 0} \|_2, \label{eq:system_equations_geometry_1} \\
    & \| \mathbf{p}_\mathrm{panel}^p - \mathbf{p}_{\rm c}^{i, 0} \|_2 \sin (\hat{\theta}_{\mathrm{S},i}^p)  = (\mathbf{p}_\mathrm{panel}^p - \mathbf{p}_{\rm c}^{i, 0})^\mathsf{T} \mathbf{n}_{\rm ULA}^p, \label{eq:system_equations_geometry_2}
\end{align}
By defining $d_p^i = \| \mathbf{p}_\mathrm{panel}^p - \mathbf{p}_{\rm c}^{i, 0} \|_2$, a linear equation system can be constructed as given by~\eqref{eq:system_linear_equations_geometry_1} and~\eqref{eq:system_linear_equations_geometry_2} (the derivation is detailed in Appendix~\ref{app:linear_system_derivation}), based on which a closed-form solution for $\mathbf{p}_{\rm c}^{i,0}$ and $d_{p_\mathrm{r}}^i$ can be obtained by using a least-squares approach. Given the initial $\mathbf{p}_{\rm c}^{i,0}$, the initial $\epsilon_i^0$ can be obtained as $\epsilon_i^0 = \frac{1}{|\mathcal{P}_{\rm LoS}(i)|} \sum_{p \in \mathcal{P}_{\rm LoS}(i)} (\hat{\tau}_{\mathrm{S},i}^p - \| \mathbf{p}_\mathrm{panel}^p - \mathbf{p}_{\rm c}^{i,0} \|_2/c)$.


\begin{figure*}[t!]
    \begin{align}
        (\mathbf{n}_{\rm ULA}^p)^\mathsf{T} \mathbf{p}_{\rm c}^{i,0} + \sin (\hat{\theta}_{\mathrm{S},i}^p) \cdot d_{p_\mathrm{r}}^i & = (\mathbf{n}_{\rm ULA}^p)^\mathsf{T} \mathbf{p}_\mathrm{panel}^p - c \cdot \hat{\tau}_{\mathrm{TDoA},i}^p \sin (\hat{\theta}_{\mathrm{S},i}^p), \quad \forall p \in \mathcal{P}_{\rm LoS}(i), \label{eq:system_linear_equations_geometry_1} \\
        2 (\mathbf{p}_\mathrm{panel}^p - \mathbf{p}_\mathrm{panel}^{p_\mathrm{r}})^\top \mathbf{p}_{\rm c}^{i,0} + 2 c \cdot \hat{\tau}_{\mathrm{TDoA},i}^p d_{p_\mathrm{r}}^i & = \|\mathbf{p}_\mathrm{panel}^p\|_2^2 - \|\mathbf{p}_\mathrm{panel}^{p_\mathrm{r}}\|_2^2 - (c \cdot \hat{\tau}_{\mathrm{TDoA},i}^p)^2, \quad \forall p \in \mathcal{P}_{\rm LoS}(i) \setminus p_\mathrm{r}. \label{eq:system_linear_equations_geometry_2}
    \end{align}
    \vspace{-2mm}
    \vspace{-7mm}
\end{figure*} 

\subsection{CC-based Multi-Panel Localization}
\label{subsec:cc_localization}
When user $i$ is identified as a single/non-\ac{los} user, a \ac{nn} model is adopted to map the \ac{csi} to a position. In order to extract better features, \ac{csi} pre-processing is conducted, where the antenna-frequency domain \ac{csi} $\mathbf{H}_i^p$ is first transformed to the angle-delay domain, denoted by $\bar{\mathbf{H}}_i^p$, and then it is truncated by taking the columns that have delays within the minimum \ac{toa} $\tau_{\min}$ and the maximum \ac{toa} $\tau_{\max}$ (further extended by $\epsilon_{\max}$ to account for the clock bias), as given by
\begin{equation}
    \bar{\mathbf{H}}_i^{p,\rm t} = [\bar{\mathbf{H}}_i^p]_{:,\left \lfloor{(\tau_{\min} - \epsilon_{\max}) \Delta_f N_c}\right \rfloor :\left \lceil (\tau_{\max} + \epsilon_{\max}) \Delta_f N_c\right \rceil }.
\end{equation}
The element-wise amplitude and phase of $\bar{\mathbf{H}}_i^{p,\rm t}$ are collected to generate the input feature vector $\mathbf{s}_i^p = {\rm vec}\left( \left[\log (\bar{\mathbf{H}}_i^{p,\rm t}), \angle \bar{\mathbf{H}}_i^{p,\rm t} \right] \right)$ for panel $p$. By concatenating the feature vectors from all panels, we form the final input $\mathbf{s}_i = f_{\mathrm{extr}}^\mathrm{C}(\boldsymbol{\mathcal{H}}_i) = \left[ ({\mathbf{s}_i^1})^\mathsf{T}, ({\mathbf{s}_i^2})^\mathsf{T}, \dots, ({\mathbf{s}_i^P})^\mathsf{T} \right]^\mathsf{T}$, based on which the position estimate can be obtained through $\hat{\mathbf{p}}_i = f_{\Theta}[\mathbf{s}_i]$.

\subsubsection{Loss for CC}
The general unsupervised \ac{cc} objective follows the learning principles introduced in UNILocPro~\cite{Unif_Pro_Yuha_2025}. In particular, the multi-component loss is inherited from UNILocPro, i.e., the overall loss function is
\begin{equation}\label{eq:overall_loss}
    \mathcal{L} = \omega_{\rm PD} \mathcal{L}_{\rm PD} + \omega_{\rm Tri} \mathcal{L}_{\rm Tri} + \omega_{\mathrm{M-LoS}} \mathcal{L}_{\mathrm{M-LoS}} + \omega_{\mathrm{OT}} \mathcal{L}_{\mathrm{OT}},
\end{equation}
where the pairwise distance loss $\mathcal{L}_{\rm PD}$ preserves the distance $d_{i,j} = d(\boldsymbol{\mathcal{H}}_i, \boldsymbol{\mathcal{H}}_j)$ between pairs of \ac{csi} samples; the triplet loss $\mathcal{L}_{\rm Tri}$ encourages the distance between an anchor and a positive sample to be smaller than the distance between the anchor and a negative sample; the multi-\ac{los} loss $\mathcal{L}_{\mathrm{M-LoS}}$ aligns the network outputs with the model-based estimates (in Sec.~\ref{sec:model_based_localization}) for identified multi-\ac{los}, allowing the learned channel chart to be anchored in global coordinates; and the \ac{ot} loss $\mathcal{L}_{\mathrm{OT}}$ is the map-assisted loss that aligns the estimated positions of identified single/non-\ac{los} users with the single/non-\ac{los} region $\mathcal{R}_{\mathrm{nb}}^{\mathrm{MP}}$. However, there are some different aspects: the CSI is constructed jointly from multiple spatially separated panels; the dissimilarity metric $d(\boldsymbol{\mathcal{H}}_i, \boldsymbol{\mathcal{H}}_j)$ is redesigned to account for the multi-panel architecture and clock asynchronism; and the training process includes a data augmentation stage that is specific to the asynchronous setting.


\subsubsection{Dissimilarity Metric with Clock Bias}
As a simple approach to obtain $d(\boldsymbol{\mathcal{H}}_i, \boldsymbol{\mathcal{H}}_j)$, we first compute a per-panel dissimilarity $d_p(\mathbf{H}_i^p, \mathbf{H}_j^p)$ and then fuse them, e.g., $d(\boldsymbol{\mathcal{H}}_i, \boldsymbol{\mathcal{H}}_j) = \frac{1}{P} \sum_{p=1}^{P} d_p(\mathbf{H}_i^p, \mathbf{H}_j^p)$ by averaging. For the per-panel dissimilarity $d_p(\mathbf{H}_i^p, \mathbf{H}_j^p)$, we adopt the fused \ac{gospa} and Wasserstein metric as in~\cite{Unif_Pro_Yuha_2025}, where the signature of each \ac{csi} $\mathbf{H}_i^p$ is constructed as a set of points $\{ \hat{\mathbf{p}}_{i,l}^p \}_{l = 1,2,\ldots,\hat{L}_i^p}$ and $\hat{\mathbf{p}}_{i,l}^p$ is an estimated position of a channel path, i.e., $\hat{\mathbf{p}}_{i,l}^p$ is computed from the estimated \ac{toa} $\hat{\tau}_{l,i}^p$ and \ac{aoa} $\hat{\theta}_{l,i}^p$ according to the geometric relationships in~\eqref{eq:geometric_relationships_ULA_1} and~\eqref{eq:geometric_relationships_ULA_2}.

However, the signature directly obtained from the estimated \ac{toa} $\hat{\tau}_{l,i}^p$, which is corrupted by the clock bias $\epsilon_i$, may not be reliable for computing the dissimilarity, thus leading to degraded performance. Therefore, we perform a clock bias mitigation to the estimated \ac{toa} as a pre-processing step before computing the signature, i.e., subtracting the estimated clock bias $\hat{\epsilon}_i$ (or a proxy) from $\hat{\tau}_{l,i}^p$ to obtain a compensated \ac{toa} $\hat{\tau}_{l,i}^p - \hat{\epsilon}_i$. The acquisition of $\hat{\epsilon}_i$ can be conducted in different ways, depending on the user identification:
\begin{itemize}
    \item For identified multi-\ac{los} users, the estimated clock bias $\hat{\epsilon}_i$ can be obtained from the model-based localization, as presented in Sec.~\ref{sec:model_based_localization};
    \item For identified single/non-\ac{los} users, the clock bias cannot be reliably recovered due to the lack of sufficient \ac{los} paths. Therefore, we adopt a clock-shift proxy that aims to reduce the \ac{toa} corruption. Specifically, $\hat{\epsilon}_i$ is obtained by solving a relaxed problem similar to~\eqref{eq:gradient_based_optimization}, where all panels are included, i.e., by setting $\mathcal{P}_{\rm LoS}(i) = \{ 1, 2, \ldots, P \}$. The resulting $\hat{\epsilon}_i$ should be regarded as a clock-shift alignment parameter rather than an accurate clock estimation. This is motivated by the fact that the user-specific clock bias is identical to all panels and all multi-path components. By estimating $\hat{\epsilon}_i$ through the alignment of the entire channel response, rather than individual \ac{los} paths, the impact of clock bias on the subsequent \ac{cc} dissimilarity metric can be mitigated. Note that in this case, the initial point $(\mathbf{p}_{\rm c}^{i,0}, \epsilon_i^0)$ is generated from a coarse grid search over the single/non-\ac{los} region $\mathcal{R}_{\mathrm{nb}}^{\mathrm{MP}}$ and $ [-\epsilon_{\max}, \epsilon_{\max}]$ instead of multi-lateration.
\end{itemize}

\subsubsection{Data Augmentation and Training Strategy}
Due to clock bias and its possible drift, \ac{csi} samples (for the same or different users) collected from the same physical location may exhibit different input features, which could degrade generalization performance. Therefore, we aim to make the \ac{nn} model agnostic to clock bias by learning a mapping from bias-invariant \ac{csi} patterns rather than unreliable \ac{toa} values. Specifically, we perform a two stage training process:
\begin{itemize}
    \item In the \ac{cc} training stage, given the dataset $\{ \boldsymbol{(\mathcal{H}}_i, t_i) \}_{i}$, where $t_i$ is the timestamp, the input features $\{\mathbf{s}_i\}_i$, the dissimilarity metric $\{d(\boldsymbol{\mathcal{H}}_i, \boldsymbol{\mathcal{H}}_j)\}_{i,j}$, and the model-based estimates for identified multi-\ac{los} users are first computed, and then the \ac{nn} model is trained using the multi-component loss~\eqref{eq:overall_loss} through backpropagation (the obtained parameters are denoted as $\Theta_0$).
    \item In the data augmentation stage, the main focus is to improve the generalization performance of the \ac{nn} model by augmenting the training dataset with additional samples that simulate the effect of clock bias. Specifically, for each sample $\boldsymbol{\mathcal{H}}_i$, according to the channel model in Sec.~\ref{sec:channel_model}, we generate multiple augmented samples by adding random clock bias perturbations $\boldsymbol{\mathcal{H}}_i^{\mathrm{aug},k} = \{ \mathbf{H}_i^p \odot [\mathbf{1} \cdot \mathbf{b}^\mathsf{T}(\delta_i^k)] \}_{p=1}^P$, $k = 1, 2, \dots, K$, where $K$ is the number of augmented samples; $\delta_i^k \in [-\epsilon_{\max}, \epsilon_{\max}]$ is a random variable; $\mathbf{1}$ is an all-ones column vector and $\odot$ is the Hadamard product. Then, given the trained model $f_{\Theta_0}(\cdot)$, we obtain the pseudo-label for the augmented samples as $\hat{\mathbf{p}}_i^{\mathrm{aug}} = f_{\Theta_0}\left[f_{\rm extr}^\mathrm{C}(\boldsymbol{\mathcal{H}}_i)\right]$. Therefore, the augmented dataset can be constructed as $\{ (\boldsymbol{\mathcal{H}}_i^{\mathrm{aug},k}, \hat{\mathbf{p}}_i^{\mathrm{aug}}) \}_{i,k}$, which is used to further train $f_{\Theta_0}(\cdot)$ with a loss of $\mathcal{L}_{\mathrm{MSE}} = \frac{1}{N_u K} \sum_{i=1}^{N_u} \sum_{k=1}^K \| f_{\Theta}[f_{\rm extr}^\mathrm{C}(\boldsymbol{\mathcal{H}}_i^{\mathrm{aug},k})] - \hat{\mathbf{p}}_i^{\mathrm{aug}} \|_2^2$.
\end{itemize}

\section{Numerical Results}
In this section, we numerically evaluate the performance of the proposed framework UNILocMP in a street canyon scenario (the same as~\cite{Unif_Pro_Yuha_2025}), where four antenna panels are uniformly arranged on the rooftop of a building with a $7 \, \mathrm{m}$ inter-panel spacing. The geometric center of the deployment is located at $[0,-9,57]^{\mathsf{T}}$.

\subsection{System Parameters and Dataset Generation}
The carrier frequency is chosen as $10\, {\rm GHz}$, and $50\, {\rm MHz}$ bandwidth is used with a subcarrier spacing of $120\, {\rm kHz}$. Each antenna panel is equipped with a \ac{ula} of $M_p = 64$ antennas, and the inter-antenna distance is set as half the wavelength. The user positions are distributed in the first quadrant of the map (around 1800 user positions are generated independently for both training and testing datasets), and the height of each user position is set as $1.5\, {\rm m}$. For each user position, the \ac{csi} $\boldsymbol{\mathcal{H}}_i$ is generated by Sionna RT synthetically~\cite{sionna}, where ray-tracing techniques are used to generate the propagation paths, based on which the \ac{csi} is computed. As an initial evaluation, we assume that $I(\cdot)$ is ideal, i.e., $p_\mathrm{I} = 1$, to decouple the proposed localization framework from the performance of the external \ac{los}/\ac{nlos} identification and establish a performance upper bound.\footnote{In future work, real \ac{los}/\ac{nlos} identification methods with classification errors, i.e., $p_\mathrm{I} < 1$, which can propagate to the localization, will be incorporated to assess the robustness and performance of the proposed UNILocMP under realistic scenarios.} Following the same \ac{nn} architecture as in~\cite{Unif_Pro_Yuha_2025,Angl_Prof_Step_2024}, a \ac{mlp} is adopted for $f_{\Theta}(\cdot)$. All other system parameters and hyper-parameters for different components in the framework are set as in~\cite{Unif_Pro_Yuha_2025}.

\subsection{Localization Performance Evaluation}
\begin{table}[t!]
\caption{The MAE (m) for different localization methods in multi-panel MIMO scenario (testing dataset).}
\label{tab:localization_performance_multi_panel}
\centering
\renewcommand{\arraystretch}{1.1} 
\begin{tabular}{|l|ccc|}
\hline
 & multi-LoS & single/non-LoS & all \\
\hline
\multicolumn{4}{|l|}{\textbf{under perfect synchronization} ($\epsilon_{\max} = 0$)} \\
\hline
model-based method & 0.64 & 13.87 & 9.75 \\
CC-based method & 27.91 & 24.38 & 25.47 \\
UNILocMP  & 0.64 & 3.16 & 2.38 \\
UNILocMP (aug) & \textbf{0.64} & \textbf{2.96} & \textbf{2.24} \\
\hline
fingerprinting & 1.38 & 1.37 & 1.37 \\
fingerprinting (aug) & \textbf{1.23} & \textbf{1.14} & \textbf{1.18} \\
\hline
\multicolumn{4}{|l|}{\textbf{under imperfect synchronization} ($\epsilon_{\max} = 100\, \mathrm{ns}$)} \\
\hline
model-based method & 1.34 & 21.42 & 15.16 \\
CC-based method & 31.18 & 27.11 & 28.38 \\
UNILocMP & 1.34 & 5.19 & 3.99 \\
UNILocMP (aug) & \textbf{1.34} & \textbf{4.87} & \textbf{3.77} \\
\hline
fingerprinting & 3.12 & 3.05 & 3.07 \\
fingerprinting (aug) & \textbf{1.51} & \textbf{1.34} & \textbf{1.41} \\
\hline
\end{tabular}
\vspace{-3mm}
\end{table}

In addition to UNILocMP, model-based, fingerprinting, and \ac{cc}-based localization methods are considered as benchmarks. For the model-based method, the optimization problem in~\eqref{eq:gradient_based_optimization} is solved for position estimation. For fingerprinting (with the same \ac{csi} pre-processing as UNILocMP), a \ac{nn} model is trained with a supervised learning approach using the ground-truth positions as labels. For the \ac{cc}-based method, the training of the \ac{nn} model is conducted using only $\mathcal{L}_{\rm PD}$ and $\mathcal{L}_{\rm Tri}$ in~
\eqref{eq:overall_loss}, where the dissimilarity metric is computed based on the angle delay profile of \ac{csi}~\cite{Angl_Prof_Step_2024}. Moreover, an affine transformation is applied to the output of the \ac{cc}-based method to anchor them in global coordinates, where the transformation parameters are obtained by solving a least-squares problem based on the ground-truth positions.

The performance of different methods under both perfect synchronization ($\epsilon_{\max} = 0$) and imperfect synchronization ($\epsilon_{\max} = 100\, \mathrm{ns}$) is evaluated, where the imperfect synchronization is emulated by adding a random clock bias $\epsilon_i$ uniformly distributed in $[-\epsilon_{\max}, \epsilon_{\max}]$ for each user. The \ac{mae} results are summarized in Table~\ref{tab:localization_performance_multi_panel} and some representative \ac{cdf} curves are shown in Fig.~\ref{fig:cdf_com_SingleBS_MultiPanel}. It can be observed that the proposed UNILocMP outperforms the model-based and \ac{cc}-based methods significantly under both perfect and imperfect synchronization. However, UNILocMP still exhibits a noticeable yet acceptable performance gap compared to fingerprinting, which is expected since fingerprinting leverages supervised learning with ground-truth labels, in contrast to the unsupervised training of UNILocMP. It also can be seen that the data augmentation contributes to improving the performance of UNILocMP and fingerprinting, especially under imperfect synchronization, by enhancing the generalization ability of the \ac{nn} model across different clock biases. 

\subsection{Comparison with Single-BS Localization}
\begin{table}[t!]
\caption{The MAE (m) for different localization methods in single-BS scenario (testing dataset).}
\label{tab:localization_performance_single_bs}
\centering
\renewcommand{\arraystretch}{1.1} 
\begin{tabular}{|l|ccc|}
\hline
 & \ac{los} & \ac{nlos} & all \\
\hline
\multicolumn{4}{|l|}{\textbf{under perfect synchronization} ($\epsilon_{\max} = 0$)} \\
\hline
model-based method & 0.34 & 13.79 & 9.81 \\
UNILocPro & 0.34 & 4.05 & 2.95 \\
UNILocPro (aug) & 0.34 & 3.72 & 2.72 \\
fingerprinting & 2.26 & 2.00 & 2.07 \\
fingerprinting (aug) & 1.50 & 1.54 & 1.53 \\
\hline
\multicolumn{4}{|l|}{\textbf{under imperfect synchronization} ($\epsilon_{\max} = 100\, \mathrm{ns}$)} \\
\hline
model-based method & 18.38 & 25.03 & 23.06 \\
UNILocPro & 18.38 & 8.88 & 11.69 \\
UNILocPro (aug) & 18.38 & 8.37 & 11.33 \\
fingerprinting & 3.87 & 4.24 & 4.13 \\
fingerprinting (aug) & 1.69 & 1.90 & 1.84 \\
\hline
\end{tabular}
\vspace{-3mm}
\end{table}

\begin{figure}[t]
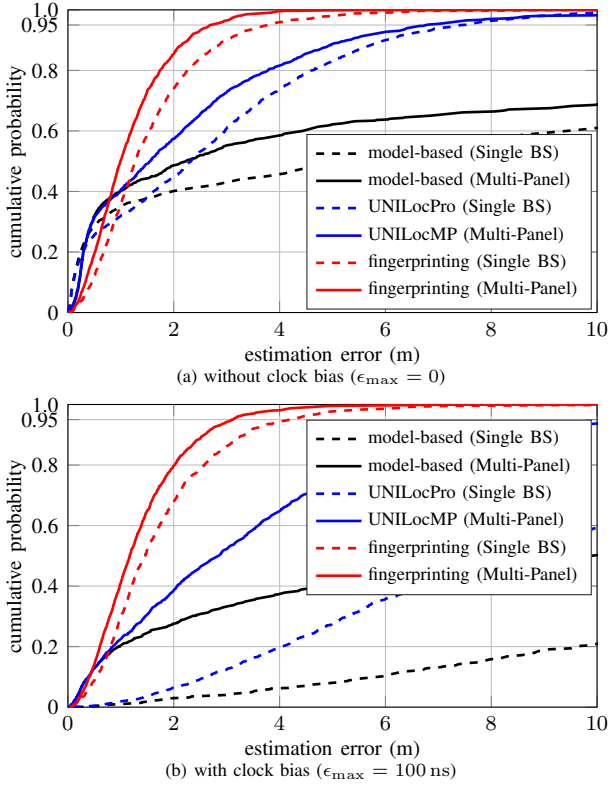

    \centering
    \begin{subfigure}{0.45\textwidth}
        \input{Figures/fig_cdf_com_SingleBS_MultiPanel_nobias.tex}
        \vspace{-3mm}
        \caption{without clock bias ($\epsilon_{\max} = 0$)}\label{fig:cdf_com_SingleBS_MultiPanel_nobias}
    \end{subfigure}
    \begin{subfigure}{0.45\textwidth}
        \input{Figures/fig_cdf_com_SingleBS_MultiPanel_bias.tex}
        \vspace{-3mm}
        \caption{with clock bias ($\epsilon_{\max} = 100\, \mathrm{ns}$)}\label{fig:cdf_com_SingleBS_MultiPanel_bias}
    \end{subfigure}
    \caption{The \ac{cdf} w.r.t.~the positioning error for different methods, where data augmentation is applied for \ac{nn}-based methods (testing dataset).}
    \label{fig:cdf_com_SingleBS_MultiPanel}
    \vspace{-3mm}
\end{figure}

To ensure a fair comparison, the single-\ac{bs} scenario employs the same total number of antennas as the multi-panel architecture. Specifically, a single \ac{ula} with $M=256$ antennas is deployed on the rooftop of the same building, with its array center located at $[0,-9,57]^{\mathsf{T}}$. The \ac{mae} for different methods for single-\ac{bs} localization is shown in Table~\ref{tab:localization_performance_single_bs}. It can be observed that the performance of the model-based method and UNILocPro~\cite{Unif_Pro_Yuha_2025} degrades significantly under imperfect synchronization, since they rely heavily on the \ac{toa} features that are corrupted by clock bias. Although data augmentation can mitigate the degradation to some extent, the improvement is very limited. On the contrary, being a purely data-driven method that learns a mapping from \ac{csi} to position, fingerprinting is less sensitive to clock bias since it does not explicitly rely on the absolute \ac{toa} values; however, it requires ground-truth labels for training.

The \ac{cdf} curves for single-\ac{bs} and multi-panel localization are shown in Fig.~\ref{fig:cdf_com_SingleBS_MultiPanel}, where UNILocPro suffers from significant performance degradation under imperfect synchronization, while UNILocMP maintains much better performance, demonstrating the robustness of UNILocMP to clock bias. Actually, all methods for multi-panel localization achieve better performance than their single-\ac{bs} counterparts, especially under imperfect synchronization, which demonstrates the benefits of leveraging multi-panel \ac{csi} for localization. In particular, multiple panels can provide more diverse and complementary geometric relationships and features, which can enhance positioning accuracy, as shown in Fig.~\ref{fig:cdf_com_SingleBS_MultiPanel_nobias}. Furthermore, they can also improve robustness to clock bias by exploiting the \ac{csi} across panels, as shown in Fig.~\ref{fig:cdf_com_SingleBS_MultiPanel_bias}. It is also worth noting that the achievable gain is dependent on deployment-specific factors such as panel placement and visibility conditions.

\section{Conclusion}
In this paper, we propose UNILocMP, a unified localization framework for multi-panel massive \ac{mimo} under clock asynchronism. By combining model-based method for multi-\ac{los} users and unsupervised \ac{cc} for single/non-\ac{los} users, together with clock-bias compensation and data augmentation, UNILocMP achieves robust localization without requiring ground-truth labels. Numerical results demonstrate superior improvements over model-based and \ac{cc}-based methods, while approaching the performance of fully-supervised fingerprinting under both perfect and imperfect synchronization. Furthermore, the comparison with single-\ac{bs} localization indicates that multi-panel architectures can provide additional geometric diversity and robustness against clock asynchronism. A systematic evaluation under real \ac{los}/\ac{nlos} identification and measured dataset is an important direction for future work.

\begin{appendices}
\section{}\label{app:linear_system_derivation}
Given $d_p^i = \| \mathbf{p}_\mathrm{panel}^p - \mathbf{p}_{\rm c}^{i,0} \|_2$, the geometric relationships in~\eqref{eq:system_equations_geometry_1} and ~\eqref{eq:system_equations_geometry_2} can be rewritten as
\begin{equation}\label{eq:tdoa_rewritten}
    c \cdot \hat{\tau}_{\mathrm{TDoA},i}^p = d_p^i - d_{p_\mathrm{r}}^i, \!\!\!\! \quad \forall p \in \mathcal{P}_{\rm LoS}(i),
\end{equation}
\begin{equation}\label{eq:aoa_rewritten}
     \sin (\hat{\theta}_{\mathrm{S},i}^p) \cdot d_p^i= (\mathbf{n}_{\rm ULA}^p)^\mathsf{T} (\mathbf{p}_\mathrm{panel}^p - \mathbf{p}_{\rm c}^{i,0}), \!\!\!\! \quad \forall p \in \mathcal{P}_{\rm LoS}(i),
\end{equation}
based on which by substituting~\eqref{eq:tdoa_rewritten} into~\eqref{eq:aoa_rewritten}, we can eliminate $d_p^i$ and obtain a linear equation, as given by
\begin{equation}\label{eq:linearized_system_1}
    \sin (\hat{\theta}_{\mathrm{S},i}^p) \cdot (d_{p_\mathrm{r}}^i + c \cdot \hat{\tau}_{\mathrm{TDoA},i}^p) = (\mathbf{n}_{\rm ULA}^p)^\mathsf{T} (\mathbf{p}_\mathrm{panel}^p - \mathbf{p}_{\rm c}^{i,0}).
\end{equation}

Moreover, from~\eqref{eq:tdoa_rewritten} we have $c \cdot \hat{\tau}_{\mathrm{TDoA},i}^p + d_{p_\mathrm{r}}^i = \| \mathbf{p}_\mathrm{panel}^p - \mathbf{p}_{\rm c}^{i,0} \|_2$, $\forall p \in \mathcal{P}_{\rm LoS}(i)$, which can be rewritten as
\begin{equation}\nonumber
(c \cdot \hat{\tau}_{\mathrm{TDoA},i}^p + d_{p_\mathrm{r}}^i)^2
= \|\mathbf{p}_\mathrm{panel}^p\|_2^2 + \|\mathbf{p}_{\rm c}^{i,0} \|_2^2 - 2 (\mathbf{p}_{\rm c}^{i,0})^\top \mathbf{p}_\mathrm{panel}^p.
\end{equation}
Then, by subtracting it of the reference panel $p_\mathrm{r}$ (for which $\hat{\tau}_{\mathrm{TDoA},i}^{p_\mathrm{r}} = 0$) from that of any other panel $p \in \mathcal{P}_{\rm LoS}(i) \setminus p_\mathrm{r}$, we can eliminate the term $\|\mathbf{p}_{\rm c}^{i,0} \|_2^2$ and obtain a linear equation:
\begin{equation}\label{eq:linearized_system_2}
    \begin{aligned}
        2 (\mathbf{p}_\mathrm{panel}^p & - \mathbf{p}_\mathrm{panel}^{p_\mathrm{r}})^\top \mathbf{p}_{\rm c}^{i,0} + (c \cdot \hat{\tau}_{\mathrm{TDoA},i}^p)^2 \\
        & + 2 c \cdot \hat{\tau}_{\mathrm{TDoA},i}^p d_{p_\mathrm{r}}^i = \|\mathbf{p}_\mathrm{panel}^p\|_2^2 - \|\mathbf{p}_\mathrm{panel}^{p_\mathrm{r}}\|_2^2.
    \end{aligned}
\end{equation}
Finally, \eqref{eq:system_linear_equations_geometry_1} and \eqref{eq:system_linear_equations_geometry_2} can be obtained by combining~\eqref{eq:linearized_system_1} for all panels $p \in \mathcal{P}_{\rm LoS}(i)$ and~\eqref{eq:linearized_system_2} for all panels $p \in \mathcal{P}_{\rm LoS}(i) \setminus p_\mathrm{r}$.

\end{appendices}

\bibliographystyle{IEEEtran}
\bibliography{main}

\end{document}